\documentclass[a4page, 10pt]{article}
\usepackage{amssymb,amsmath}
\usepackage{graphicx}
\usepackage{setspace}
\begin{document}

\title{Action-Angle Variables in Quantum Gravity}

\author{Jarmo M\"akel\"a\footnote{Vaasa University of Applied Sciences,
Wolffintie 30, 65200 Vaasa, Finland, email: jarmo.makela@vamk.fi}}

\maketitle

\begin{abstract}

We formulate an argument, based on the use of the action-angle variables and the Bohr-Sommerfeld quantization rule, to the effect that if there exists the smallest possible non-zero area, there also exists, for massive particles, the largest possible observable speed, which is a bit less than the speed of light.

\medskip


\end{abstract}

For any periodic system with $m$ degrees of freedom and completely separable characteristic function the action variables $J_k$ $(k = 1,2,\dots,m)$ are defined as: \cite{yy}
\begin{equation}
J_k = \oint p_k\,dq_k,
\end{equation}
where the variables $q_k$ are the coordinates of the configuration space, and the variables $p_k$ the corresponding coordinates of the momentum space. In Eq. (1) we have integrated over the period of the system, assuming that the classical equations of motion are satisfied. If we take the action variables $J_k$ as the coordinates of the momentum space, then the so-called angle variables $w_k\in [0,1]$ are the corresponding coordinates of the configuration space, and there exists a canonical transformation from the phase space coordinates $(q_k, p_k)$ to the new phase space coordinates, the action-angle variables $(w_k, J_k)$. In the old quantum theory, which prevailed before the  discovery of quantum mechanics around the year 1926, one postulated the so-called Bohr-Sommerfeld quantization rule:
\begin{equation}
J_k = 2\pi n_k\hbar,
\end{equation}
where $n_k = 0,1,2,\dots$, for all $k=1,2,\dots,m$. As it is well known, the Bohr-Sommerfeld quantization rule gives,  among other things, for the one-electron atoms an energy spectrum, which essentially agrees with the energy spectrum obtained from the proper quantum-mechanical consideration. In the classical limit, where the quantum numbers $n_k$  become large, the results obtained from the Bohr-Sommerfeld quantization rule generally give pretty good approximations to the spectra of the observable quantities.

  It was suggested long ago by Bekenstein that the event horizon area $A$ of a black hole is an adiabatic invariant of the hole. Since, as it was pointed out by Bekenstein, the action variables of any periodic system are its adiabatic invariants, it is natural to think that $A$ is proportional to an action variable $J$ of the hole.  In other words, \cite{kaa}, \cite{koo}
\begin{equation}
A = \alpha J,
\end{equation}
where $\alpha$ is an appropriate constant. If this is the case, the Bohr-Sommerfeld quantization rule implies that the eigenvalues of the area $A$ are of the form:
\begin{equation}
A_n =2\pi n\alpha\hbar,
\end{equation}
where $n=0,1,2,\dots$. In other words, the event horizon area of the black hole has a {\it discrete spectrum with an equal spacing}. Such spectra for the black hole event horizon area have been proposed by several authors on various grounds. \cite{nee} In most approaches the area eigenvalues have been taken to be:
\begin{equation}
A_n = 4n\ell_{Pl}^2\ln(2),
\end{equation}
since this is consistent with the Bekenstein-Hawking entropy law. \cite{vii} In Eq. (5) $\ell_{Pl} :=\sqrt{\frac{\hbar G}{c^3}}\approx 1.6\times 10^{-35}m$ is the Planck length. {\footnote {In loop quantum gravity the area eigenvalues of an arbitrary, spacelike two-surface are of the form:
\[
A = \gamma\ell_{Pl}^2\sum_p\sqrt{j_p(j_p+1)},
\]
where $\gamma$ is an appropriate constant. When the quantum numbers $j_p = 0, 1/2, 1, 3/2,\dots$ are very large, the area spectrum agrees with that in Eq. (5). A modification of loop quantum gravity, which produces Eq. (5) explicitly, has been formulated by Krasnov. \cite{kootoo}}}

  In this paper we shall take Eq. (3) as a starting point, and see, where it will take us. Unless otherwise stated, we shall always use the natural units, where $\hbar = G = c = 1$. Eqs. (4) and (5) imply that in the natural units we may write:
\begin{equation}
A = \frac{2\ln(2)}{\pi}J.
\end{equation}

   It has been shown in several papers that from the point of view of an observer with constant proper acceleration $a$, just outside of the event horizon of a black hole we may take
\begin{equation}
H = \frac{a}{8\pi}A
\end{equation}
as the Hamiltonian of the hole. \cite{kuu} In Ref. \cite{seite}, for instance, Eq. (7) was obtained by means of an analysis of the properties of the Brown-York energy of the gravitational field just outside of the horizon of spacetime, whereas in Ref. \cite{kasi} it was found by means of a systematic study of the Arnowitt-Deser-Misner (ADM) formulation of general relativity. Eq. (6) implies that we may write the Hamiltonian as a function of the action variable $J$ as:
\begin{equation}
H(J) = \frac{\ln(2)}{4\pi^2}aJ,
\end{equation}
and hence the corresponding angle variable $w$ has the property:
\begin{equation}
\dot{w} = \frac{\partial H(J)}{\partial J} = \frac{\ln(2)}{4\pi^2}a,
 \end{equation}
where the dot means the derivative with respect to the proper time $\tau$ of the observer. Since the proper acceleration $a$ is assumed to be a constant, no matter what may happen to the black hole, we must have:
\begin{equation}
w = \frac{\ln(2)}{4\pi^2}a\tau + C,
\end{equation}
where $C$ is a constant. 

    To understand, what Eq. (10) means, we must recall the basic properties of the Rindler spacetime. As it is well known, Rindler spacetime is just the ordinary, flat, two-dimensional Minkowski spacetime equipped with the Rindler coordinates $(t,x)$, which are related to the flat Minkowski coordinates $(T,X)$ such that: \cite{ysi}
\begin{subequations}
\begin{eqnarray}
T &=& x\sinh(t),\\
X &=& x\cosh(t).
\end{eqnarray}
\end{subequations}
By means of the Rindler coordinates we may write the line element of flat spacetime as:
\begin{equation}
ds^2 = -x^2\,dt^2 + dx^2.
\end{equation}
For an observer with constant spatial coordinate $x$ the proper accelereration $a$ is a constant, and it is related to $x$ as:
\begin{equation}
a = \frac{1}{x}.
\end{equation}
For such observer the temporal coordinate $t$ gives the {\it boost angle}. The velocity of the observer is
\begin{equation}
v = \tanh(t).
\end{equation}
The Rindler spacetime has the {\it Rindler horizon}, where $x=0$. At the Rindler horizon
\begin{equation}
T = \pm X.
\end{equation}

     The Rindler spacetime may be used in the study of the properties of any spacetime close to its horizon. As a four-dimensional generalization of the Rindler spacetime we  may consider spacetime equipped with the line element
\begin{equation}
ds^2 = -y\,dt^2 + \frac{(y')^2}{4y}\,d\lambda^2 + h_{jk}\,d\chi^j\,d\chi^k.
\end{equation}
When obtaining this line element from the line element (12) we have first defined $y:=x^2$, and then assumed that $y$ is a function of the new spatial coordinate $\lambda$. $h_{jk}$ is the metric tensor induced on the spacelike two-surface, where $\lambda = constant$, and it is assumed to be a function of $\lambda$, and of the coordinates $\chi^j$ $(j = 2,3)$ on this two-surface. The prime means the derivative with respect to $\lambda$. 

  The line element (16) approximates the line element of almost any spacetime just outside of its horizon. It was used in Refs. \cite{seite} and \cite{kasi}, when obtaining Eq. (7). For an observer with constant proper acceleration the spatial coordinates $\lambda$ and $\chi^j$ are constants, and we have:
\begin{equation}
a = \frac{1}{\sqrt{y}}.
\end{equation}
Again, the time coordinate $t$ is the boost angle of the observer.

  Eqs. (16) and (17) enable us to understand the meaning of Eq. (10). For an observer with constant spatial coordinates $\lambda$ and $\chi^j$ the proper time elapsed during the lapse $dt$ of the Rindler time $t$ is:
\begin{equation}
d\tau = \sqrt{y}\,dt = \frac{1}{a}\,dt,
\end{equation}
and since the proper acceleration $a$ is assumed to be a constant, Eq. (10) implies:
\begin{equation}
w = \frac{\ln(2)}{4\pi^2}t + C.
\end{equation}
Hence we observe that the angle variable $w$ agrees, up to multiplicative and additive constants, with the {\it boost angle} of the observer. In other words, the horizon area of the black hole and the boost angle of the observer constitute, up to constants, a canonically conjugate pair of variables.

   At this point we recall that:
\begin{equation}
0 \le w \le 1.
\end{equation}
Eq. (19) implies that the boost angle of the observer has {\it fixed bounds}. For symmetry reasons we choose $C=\frac{1}{2}$ in Eq. (19). With this choice we find, denoting the boost angle of the observer by $\phi$:
\begin{equation}
-\frac{2\pi^2}{\ln(2)} \le \phi \le \frac{2\pi^2}{\ln(2)}.
\end{equation}

   The result that the boost angle of the observer has specific bounds, instead of extending from the negative to the positive infinity, is most remarkable. It was a simple consequence of the Bohr-Sommerfeld quantization rule, and an idea, first proposed by Bekenstein, that the event horizon area of the black hole is proportional to an action variable of the system. We shall investigate the physical implications of Eq. (21) at the end of this paper. Right now, however, we pose a question of whether Eq. (21) is restricted just to black holes, or does it have more general relevance in quantum gravity.

   To begin with, we recall that with every point $P$ of four-dimensional curved spacetime we may associate a tangent space $T_P$, which is a flat, four-dimensional Minkowski spacetime. In this spacetime we define vectors ${\bf e}_T$, ${\bf e}_X$, ${\bf l}_2$ and ${\bf l}_3$ such that the position vector of an arbitrary point of $T_P$ may be written as:
\begin{equation}
{\bf r} = \tau\cosh(\phi u^1)\,{\bf e}_T + \tau\sinh(\phi u^1)\,{\bf e}_X + u^2\,{\bf l}_2 + u^3\,{\bf l}_3,
\end{equation}
where $u^1, u^2, u^3 \in [0,1]$ and the vectors ${\bf e}_T$ and ${\bf e}_X$ have the properties:
\begin{subequations}
\begin{eqnarray}
{\bf e}_T \cdot{\bf e}_T &=& -1,\\
 {\bf e}_X \cdot{\bf e}_X &=& 1,\\
{\bf e}_X\cdot{\bf e}_T &=& 0.
\end{eqnarray}
\end{subequations}
The vectors ${\bf l}_2$ and ${\bf l}_3$, in turn, are spacelike vectors orthogonal to the vectors ${\bf e}_T$ and ${\bf e}_X$. In other words:
\begin{equation}
{\bf l}_j\cdot{\bf e}_T = {\bf l}_j\cdot{\bf e}_X = 0
\end{equation}
for all $j=2,3$. We also have:
\begin{equation}
{\bf l}_2\cdot{\bf l}_3 = \frac{1}{2}(l_2^2 + l_3^2 - l_{23}^2),
\end{equation}
where we have defined:
\begin{equation}
{\bf l}_{23} := {\bf l}_2 - {\bf l}_3.
\end{equation}

  When written in terms of the coordinates $\tau$, $u^1$, $u^2$ and $u^3$, the line element of spacetime takes the form:
\begin{equation}
ds^2 = -d\tau^2 + \tau^2\phi^2\,(du^1)^2 + l_2^2\,(du^2)^2 + (l_2^2 + l_3^2 - l_{23}^2)\,du^2\,du^3 + l_3^2\,(du^3)^2.
\end{equation}
Our tangent space consists of a Cartesian product of a parallellogram with edge vectors ${\bf l}_2$ and ${\bf l}_3$, and of a flat spacetime with coordinates $\tau$ and $u^1$. The area of the parallellogram is 
\begin{equation}
A = \frac{1}{2}\sqrt{4l_2^2l_3^2 - (l_2^2 + l_3^2 - l_{23}^2)^2}.
\end{equation}
For an observer with constant coordinates $u^1, u^2, u^3 \in [0, 1]$ $\tau$ is the proper time of the observer, and the observer is in a free fall. The quantity $\phi u^1$, in turn, gives the boost angle of the observer. Making $l_2$, $l_3$ and $\tau$ small we may make the tangent space as small as we like. 

   As it is well known, there is an alternative to the standard ADM formulation of general relativity, where the components $q_{ab}$ $(a, b = 1, 2, 3)$ of the metric tensor induced on the spacelike hypersurface of spacetime, where the time coordinate $t=constant$, are used as the coordinates of the configuraton space.  In this alternative formulation the so-called {\it densitized triads}
\begin{equation}
\tilde{E}_I^a := \sqrt{q}E_I^a
\end{equation}
on that hypersurface are considered as the coordinates of the configuration space, and the quantities
\begin{equation}
p_a^I := \frac{1}{8\pi}K_a^I = \frac{1}{8\pi}E^{Ib}K_{ab}
\end{equation}
as the corresponding coordinates of the momentum space. \cite{kymppi} In Eqs. (29) and (30) the quantities $E_I^a$ $(a, I = 1,2,3)$ are the components of the {\it triads}. The triad field $E_I^a$ has the property
\begin{equation}
q_{ab}E_I^aE_J^b = \delta_{IJ}.
\end{equation}
The quantities $K_{ab}$ are the components of the exterior curvature tensor on the $t=constant$ hypersurface of spacetime. In what follows, we shall denote the indices referring to the flat Minkowski coordinates attached to the given point of spacetime by numbers equipped with bars. 

   Eq. (27) implies that we may take:
\begin{subequations}
\begin{eqnarray}
E_{\bar{1}}^1 &=& \frac{1}{\tau\phi},\\
K_{11} = -\frac{1}{2} \dot{q}_{11} &=&- \tau\phi^2.
\end{eqnarray}
\end{subequations}
So we find:
\begin{subequations}
\begin{eqnarray}
\tilde{E}_{\bar{1}}^1 &=& A,\\
p_1^{\bar{1}} &=& -\frac{1}{8\pi}\phi,
\end{eqnarray}
\end{subequations}
where we have used Eq. (28). The other components of $\tilde{E}_I^a$ and $p_a^I$ will vanish in the limit, where $\tau\rightarrow 0$. Hence we observe that the area $A$ of the parallellogram, and the boost angle $\phi$ of an observer with $u^1=1$, multiplied by $-\frac{1}{8\pi}$, constitute a canonically conjugate pair of variables. This means that the phase space in general relativity may be spanned by the areas and the boost angles measured in a local Minkowski system of coordinates by a freely falling observer at each point of spacetime.

  There are several approaches to quantum gravity, such as loop quantum gravity, where the area of an arbtrary spacelike two-surface of spacetime has a discrete spectrum. \cite{yytoo} If we assume that the area $A$ of an arbitrary spacelike two-surface is proportional to an action variable $J$ as in Eq. (6), which means that
\begin{equation}
J = \frac{\pi}{2\ln(2)}A,
\end{equation}
then the Bohr-Sommerfeld quantization rule (2) implies hat the area eigenvalues are identical to those in Eq. (5). Even more interesting, however, is that the corresponding angle variable $w$ is related to the boost angle $\phi$ of the observer as:
\begin{equation}
w = \frac{2\ln(2)}{\pi}p^{\bar{1}}_1 + C = \frac{\ln(2)}{4\pi^2}\phi + C,
\end{equation}
because the transformation from the variables $(\tilde{E}^1_{\bar{1}}, p^{\bar{1}}_1)$ to the variables $(w, J)$ is canonical. Again, if we choose $C = \frac{1}{2}$ and take into account that $0 \le w\le 1$, we find that the boost angle $\phi$ has fixed bounds, such that
\begin{equation}
-\frac{2\pi^2}{\ln(2)} \le \phi\le \frac{2\pi^2}{\ln(2)}.
\end{equation}
These are exactly the same bounds as in Eq. (21). Hence it appears that Eq. (21) is a generic result, which holds not only for observers with constant proper acceleration just outside of a horizon of spacetime, but also for observers in a free fall. 

    An obvious consequence of Eq. (36) is that there exists a certain {\it maximum velocity}, which is a bit {\it less} than the speed of light. Since between the velocity $v$ and the boost angle $\phi$ there is the relationship:
\begin{equation}
v = c\tanh(\phi),
\end{equation}
Eq. (36) implies that this maximum velocity is:
\begin{equation}
\begin{split}
v_{max} &= c\tanh\left\lbrack\frac{2\pi^2}{\ln(2)}\right\rbrack\\
             &= c\sqrt{1 - \frac{1}{\cosh^2\left\lbrack\frac{2\pi^2}{\ln(2)}\right\rbrack}}\\
             &\approx \left\lbrace 1 - \frac{1}{2\cosh^2\left\lbrack\frac{2\pi^2}{\ln(2)}\right\rbrack}\right\rbrace c\\
             &\approx(1 - 3.6\times 10^{-25})c,
\end{split}
\end{equation}
If a massive particle raced with this speed with a photon across a galaxy with diameter $10^5$ light years, it would lag less than 0.4 mm behind the photon at the end of he race. The corresponding {\it maximum energy} of a particle with mass $m$ is:
\begin{equation}
E_{max} = mc^2\cosh\left\lbrack\frac{2\pi^2}{\ln(2)}\right\rbrack.
\end{equation}
For a proton, for instance, this energy is around $1.1\times 10^{21}eV$. Interestingly, this is roughly of the same order of magnitude as has been the highest energy ever measured for the particles of the cosmic rays, or $(3.2\pm 0.9)\times 10^{20}eV$. It is still uncertain, what this so-called "Oh-My-God-particle" really was, but it is believed to have been a proton.   \cite{kaatoo}

   A natural interpretation of our results is that it is impossible for an observer to detect massive particles with velocities greater than $v_{max}$ in Eq. (38) or with energies greater than $E_{max}$ in Eq. (39). In a nutshell, the content of this paper may be summarized such that if there exists the smallest possible area, then there also exists the largest possible speed, which is a bit less than the speed of light. When drawing this conclusion we used an argument based on Bekenstein's idea that the black hole event horizon area is proportional to an action variable of the system. It was found that the area $A$ and the boost angle $\phi$ of an observer with constant proper acceleration just outside the horizon, multiplied by $-\frac{1}{8\pi}$, constitute a canonical pair of variables. Since the angle variable $w\in [0,1]$ and the action variable $J$ constitute a canonical pair of variables, $\phi$ equals with $w$ up to multiplicative and additive constants, and has therefore fixed bounds. Similar results were found for freely falling observers.

 Since the boost angle and the area are, up to a constant, canonically conjugate variables, the question about the possible Lorentz invariance of the maximum speed of massive particles is closely related the question about the Lorentz invariance of the minimum area in quantum gravity. It was suggested by Rovelli and Speziale in Ref. \cite{neetoo} that local Lorentz boosts change the probability distribution of the measured area eigenvalues of spacelike two-surfaces, but not the area eigenvalues themselves. In this sense the minimum area in quantum gravity is Lorentz invariant. It appears to the author that if one accepts the reasoning in Ref. \cite{neetoo}, one is also forced to accept the Lorentz invariance of the maximum speed. In other words, the maximum speed of massive particles is given by Eq. (38) for all observers.

\medskip

{\bf Data Availability Statement:} All data used in this paper may be found in the references listed in the paper.

\end{document}